# Managing Critical Spreadsheets in a Compliant Environment

Soheil Saadat
President and CEO
Prodiance Corporation
5000 Executive Parkway, Suite 270
San Ramon, CA 94583 – USA
soheil.saadat@prodiance.com

**Spreadsheets – The Hidden Risks**

The use of uncontrolled financial spreadsheets can expose organizations to unacceptable business and compliance risks, including errors in the financial reporting process, spreadsheet misuse and fraud, or even significant operational errors. These errors have been well documented and thoroughly researched by the European Spreadsheet Risks Interest Group [EuSpRIG, 2005]. With the advent of regulatory mandates such as SOX 404 and FDICIA in the U.S., and MiFID, Basel II and Combined Code in the UK, leading tax and audit firms are now recommending organizations automate internal controls over critical spreadsheets and other end-use computing applications, including Microsoft Access databases. At a minimum, auditors mandate version control, change control and access control for operational spreadsheets, and more advanced controls for critical financial spreadsheets [PwC, 2004]. While regulatory compliance has remained a key business driver, many organizations are implementing spreadsheet controls to manage operational risk, and to achieve sound corporate governance and process improvements.

The inherent complexities of operational and financial spreadsheets expose technological shortcomings of available spreadsheet management solutions. Specifically, financial spreadsheets often contain external links to other spreadsheets and databases. For example, a consolidated revenue spreadsheet may contain inbound links from individual product revenue reports, and outbound links providing results to executive dashboards or the overall balance sheet. Often these critical spreadsheets reside in employee desktops, in email attachments, or on corporate shared drives – an uncontrolled environment that is absent traditional IT controls. As such, security over these critical spreadsheets tends to be weak, access is often not controlled, file versioning is not implemented, there is no visibility into changes being made, nor validation that external links are correct.

The bottom line is critical business decisions are being made everyday based on data produced by critical spreadsheets, yet executives have little confidence or trust in the data being produced in uncontrolled environments.

**Spreadsheet Links Create a Technology Challenge**

Auditors and executives alike agree that the right solution to address these challenges is "to move to an automated, controlled, yet flexible technology-based environment" [Ernst & Young, 2007]. Centralizing spreadsheet control creates a new system of record for all critical spreadsheets, and enables organizations to apply auditor recommended IT controls such as versioning, security and access control, records retention, archival and backup, change control and workflow automation [PwC, 2004]. However, simply moving spreadsheets into a document management system through traditional methods often breaks the links, and this requires many additional man hours to re-establish the links. Without the use of technology, this is a labor-intensive and manual process, and the lack of visibility into spreadsheet links and lack of documentation compounds the problem. Notwithstanding, today's commercial document management systems are not designed to work seamlessly with Microsoft Office Excel to preserve and update spreadsheet links. If a spreadsheet is moved into a document management system, or even to another network file location, the links will break. Many companies have tried migrating critical spreadsheets into document management systems, only to have exasperated end users who cannot update their spreadsheet input data through the resulting broken links. These projects have failed miserably, leaving IT project managers, auditors and financial executives to look for alternatives.

**Solution for Managing Linked Spreadsheets**

Fortunately, for these technology and business challenges there is a solution. A proven approach to efficiently managing linked spreadsheets in a controlled and compliant environment:

- Automates spreadsheet discovery, documentation and risk analysis, including the creation dependency diagrams to provide visibility into existing spreadsheet links.
- Provides tools for the migration of critical spreadsheets into secure, document management repository while automatically re-establishing any and all links (to their new web folder location, e.g. http://sharepoint/).
- Generates a migration or inventory log of spreadsheets migrated and any changes to spreadsheet links.
- Incorporates a technology integration layer enabling leading document management systems (e.g. SharePoint) to automatically update real-time data feeds through spreadsheet links.
- Incorporates auditing of spreadsheet changes down to the cell level to satisfy change control requirements.
- Automates spreadsheet change request, testing, review and approval processes via workflow for both developers and end users.

**Automating Discovery, Documentation, and Risk Analysis**

To help organizations automate spreadsheet inventory efforts, discovery tools can be leveraged to search across a wide variety of data sources and report on spreadsheets and other end-user applications (including Access databases) that are being used within an organization [Protiviti, 2006]. Through a consolidated interface, users can search and generate an inventory report on spreadsheets meeting generic or custom search criteria (e.g. all spreadsheets where "Date Last Saved" equals "2006" or "2007" would represent spreadsheets last saved during the past year, or all spreadsheets where "Risk" equals "High" or "Medium").

Spreadsheet analysis tools can perform a risk-based analysis (based on complexity and materiality) while automatically generating documentation about critical spreadsheets. For example a cell and formula diagnostic report can show formulas with errors conditions, uncover very hidden worksheets, invisible cells, inconsistent formulas, and a whole host of other key areas of risk. Inventory reports can also be generated listing all critical spreadsheets (and their dependents) along with a host of documentation, including date created, date last modified, owner, location, number of external links, number of worksheets, number of formulas, and many other criteria to show complexity. Spreadsheet experts often refer to this process as a model audit (or analysis of the correctness) of spreadsheets [Croll, 2007].

**Link Migration Tools**

Once relevant spreadsheets have been discovered, they should be migrated from uncontrolled desktops and shared drives into a secure, web repository. As mentioned above, this can be a challenging and time consuming task given the alternative of manual copy/paste operations and manually reestablishing links to dependant spreadsheets. An automated and proven approach requires a migration tool that automatically updates any links based on the new repository location of the spreadsheets. By moving spreadsheets into a secure, web based document management repository, a host of features and controls such as improved security and access control, versioning, check-in/check-out, records retention, workflow automation, and file level audit trails are available.

**Support for Leading Document Management Systems**

An additional requirement includes support for leading 3$^{rd}$ party document management systems, including Microsoft Office SharePoint Services and SharePoint Server. Most organizations already have these technologies in place, and will want to leverage their investments. As such, spreadsheet control solutions should support leading document management repositories via the WebDAV protocol, which allows the repository to expose itself as a network drive letter and provides a seamless end user experience. For example, a user opening a controlled spreadsheet from a SharePoint repository could simply use the standard File > Open dialog from within Excel to open a critical spreadsheet, check it out, makes changes, and then use the standard File > Save dialog to save the changes, check the file in, and then automatically submit it into a review and approval workflow process where electronic signatures are captured. It is through WebDAV that Excel can automatically and successfully update spreadsheet links with 3$^{rd}$ party document management repositories.

With this approach, the impact to end users is minimized, as is the impact to existing business processes. In addition, having no software requirements for client computers insulates end users from the complexities of the technology, minimizes training requirements, and reduces IT support.

**Change Management**

Following auditor guidance, there are two aspects of change management that are required for critical spreadsheets, a detailed audit history of changes down to the cell level, and an automated workflow process to enforce the requesting, incorporating, reviewing and approving of all changes. Spreadsheet control solutions can capture changes to data, formulas, macros, queries, and also report on row and column insertions/deletions and alert users via email on any these changes. This provides an extensive database for management reporting. Combined with automated workflows, spreadsheets can be routed for review, validation and approval to enforce corporate change management policies and to ensure that financial spreadsheets are appropriately reviewed during the quarter and year-end close process. [Panko and Ordway, 2008]

**Business Benefits**

By incorporating the technology capabilities described in this paper, organizations can take a proactive approach to automating the spreadsheet compliance lifecycle, from discovery and inventory, to risk analysis, management and automation or key financial workflows. Whether the need is being driven by the need to satisfy regulatory compliance mandates, improve existing business processes, or to better manage operational risk, controls over critical spreadsheets can be automated to help restore confidence and trust in key financial data analyzed and reported in spreadsheets. Business benefits realized to date by organizations successfully automating spreadsheet controls include:

1. Improved visibility into end user computing environment (e.g. spreadsheets and Access databases) via management and change reports
2. Improved compliance with regulatory mandates, including SOX 404, FDICIA, MiFID, Basel II and Combined Code that satisfies auditor scrutiny
3. Improved internal controls via technology automation
4. Improved financial processes via workflow automation
5. Improved spreadsheet development and use
6. Improved productivity for end users

**References:**

[1] Ernst & Young, April 2007. *Insurance Industry Struggling to Meet Heightened Data Management Demand According to Ernst & Young Actuarial Transformation$^{(TM)}$ Roundtable*. Available online: http://www.thefreelibrary.com.

[2] EuSpRIG, 2005. Available online: http://www.eusprig.org/stories.htm 8/4/05 9:20.

[3] Croll, G. J. 2007. *A Typical Model Audit Approach: Spreadsheet Audit Methodologies in the City of London*. Available online: http://arxiv.org/ftp/arxiv/papers/0712/0712.2591.pdf.

[4] Panko, R. and Ordway, N., April 2005. *Sarbanes-Oxley: What About all the Spreadsheets?* Proceedings of the European Spreadsheet Risks Interest Group (EuSpRIG) 2005 15-47 ISBN:1-902724-16-X. Available online: http://arxiv.org/abs/0804.0797v1

[5] PricewaterhouseCoopers, July 2004. *The Use of Spreadsheets: Considerations for Section 404 of the Sarbanes-Oxley Act*. http://www.pwc.com/images/gx/eng/fs/insu/rt5.pdf Accessed 27 May 08 10:12

[6] Protiviti, February 2006. *Excel in Managing Spreadsheet Risk*. Internal Audit & Business Risk, February 2006, Page 32. http://www.protiviti.com/downloads/PRO/pro-gb/Excel_in_managing_spreadsheet_risk.pdf Accessed 27 May 08 10:08